\documentclass[
reprint,
amsmath,
amssymb,
aps,
prx,
]{revtex4-2} 

\usepackage{graphicx}
\usepackage{dcolumn}
\usepackage{bm}
\usepackage{siunitx}
\usepackage{braket}
\usepackage{xcolor}
\usepackage[normalem]{ulem}
\usepackage[utf8]{inputenc}
\usepackage{upgreek}

\usepackage{hyperref}
\hypersetup{
    colorlinks,
    linkcolor={black},
    citecolor={blue!80!black},
    urlcolor={blue!80!black}
}

\newcommand{\diff}{\text{d}}
\newcommand{\imu}{\text{i}}
\newcommand{\expu}{\text{e}}
\newcommand{\dg}[1]{{#1}^\dagger}

\newcommand{\re}[1]{\text{Re}#1}

\makeatletter
\renewcommand*{\@fnsymbol}[1]{\ensuremath{\ifcase#1\or \dagger \or *\or \ddagger\or
   \mathsection\or \mathparagraph\or \|\or **\or \dagger\dagger
   \or \ddagger\ddagger \else\@ctrerr\fi}}
\makeatother

\makeatletter
\def\clearfmfn{\let\@FMN@list\@empty}
\makeatother

\begin{document}

\title{Observation of superradiant bursts in a cascaded quantum system}

\author{Christian Liedl}
\author{Felix Tebbenjohanns}
\author{Constanze Bach}
\author{Sebastian Pucher}\thanks{Present address: Max-Planck-Institut für Quantenoptik, Hans-Kopfermann-Straße 1, 85748 Garching, Germany
}
\author{Arno Rauschenbeutel}
\author{Philipp Schneeweiss}\email{philipp.schneeweiss@hu-berlin.de}

\affiliation{Department of Physics, Humboldt-Universität zu Berlin, 10099 Berlin, Germany}

\date{\today}

\begin{abstract}
\noindent
Dicke superradiance describes the collective radiative decay of a fully inverted ensemble of two-level atoms. We experimentally investigate this effect for a chiral, i.e.,~direction-dependent light--matter coupling. Despite a fundamentally different interaction Hamiltonian which has a reduced symmetry compared to the standard Dicke case, we do observe a superradiant burst emission. The burst occurs above a threshold number of atoms, and its peak power scales faster with the number of atoms than in the case of free-space Dicke superradiance. We measure the first-order coherence of the burst emission and experimentally distinguish two regimes, one dominated by the coherence induced during the excitation process and the other governed by vacuum fluctuations. Our results shed light on the collective radiative dynamics of cascaded quantum many-body systems, i.e., a system in which each quantum emitter is only driven by light radiated by emitters that are further upstream in the cascade. 
Our findings may turn out useful for generating multi-photon Fock states as a resource for quantum technologies.
\end{abstract}

\maketitle

\paragraph*{Introduction.}

When a single quantum emitter interacts with a propagating light field, one usually assumes that the emitter--light coupling strength is independent of the sense of propagation of the light, forward or backward~\cite{haroche2006exploring}. 
However, under certain circumstances, this symmetry can be broken, rendering the interaction between the emitter and the field mode propagation direction-dependent, or ``chiral''~\cite{lodahl2017chiral}.
This chiral coupling lends itself to the implementation of spin-controlled non-reciprocal devices~\cite{sayrin2015nanophotonic} and is a powerful resource in quantum information~\cite{mahmoodian2016quantum,bechler2018passive}.
When more than one quantum emitter is chirally coupled to a common optical mode, this realizes a so-called cascaded quantum system~\cite{gardiner1993driving, carmichael1993quantum}. There, each quantum emitter is only driven by light radiated by emitters that are further upstream in the cascade. In other words, there is only one direction in which information about each sub-system can propagate through the ensemble.  

One of the hallmark collective effects of ``symmetric'' quantum optics is Dicke superradiance~\cite{dicke1954coherence, gross1982superradiance}, where an ensemble of initially excited atoms emits light in a short, so-called superradiant burst. The latter is characterized by an initial increase of the emitted optical power, which is due to a spontaneous synchronization of the initially independent atomic dipoles. The study of such superradiant emission in spatially extended atomic ensembles has seen a revival in recent years, driven by an increasing level of experimental control~\cite{skribanowitz1973observation, gross1976observation, gross1979maser, bohnet2012steady, goban2012demonstration, norcia2016superradiance, ferioli2021laser-driven} and theoretical efforts to understand the intricacies of such many-body quantum systems~\cite{ostermann2012cascaded, masson2020many, masson2022universality, sierra2022dicke, robicheaux2021theoretical, rubiesbigorda2021superradiance, lemberger2021radiation, plankensteiner2022quantumcumulants}. 
In his seminal work, Dicke analyzed an atomic ensemble which featured particle exchange symmetry, i.e., the system Hamiltonian is identical under the exchange of two participating atoms~\cite{dicke1954coherence, arecchi1972atomic}. Because of this symmetry, the ensemble remains in the sub-space of so-called symmetric Dicke states, which makes the problem analytically solvable even for a large number of atoms~\cite{gross1982superradiance}.
In contrast, the interaction Hamiltonian of a cascaded quantum system breaks the particle exchange symmetry~\cite{stannigel2012driven, pichler2015quantum, kumlin2020nonexponential}.
This implies, e.g., that the first atom in the cascade is completely independent of the rest of the ensemble, while the dynamics of the last atom depends on all other atoms.
Recently, it has been theoretically predicted that superradiant bursts  occur even for perfectly unidirectional coupling~\cite{cardenas2022many}. This dynamics has, however, not been observed experimentally.

Here, we observe the superradiant burst emission of light from about $1000$ cesium atoms that are chirally coupled to a nanophotonic waveguide.
The atomic ensemble extends over thousands of optical wavelengths and features waveguide-mediated long-range interactions~\cite{sheremet2023waveguide}. 
The atoms are excited through the waveguide by means of a short resonant optical pulse. Following this excitation, the ensemble emits a burst of light into the waveguide if the number of atoms exceeds a threshold value.
Interestingly, we find that the peak power of this burst scales faster with the number of atoms than in the case of Dicke superradiance in free space.
We show that, because of the initial coherence that is present in the atomic ensemble after excitation, the superradiant bursts are coherent with respect to the excitation laser field for a wide range of excitation pulse areas.
However, when the excitation pulse area is chosen such that the ensemble is coherently prepared close to full inversion, the coherence of the superradiant burst with respect to the excitation laser field is lost. 
This shows that the burst emitted by a fully inverted ensemble is induced by vacuum fluctuations.
Consequently, it exhibits a vanishing expectation value of the electric field, meaning that its phase is undefined. 
Still, even then, we find that the field at the beginning of the burst is coherent with the field at later times, indicating that the vacuum-induced burst is primarily emitted into a single temporal mode. 
Finally, we present a model whose computational complexity only scales linearly with the number of atoms and which accurately describes all our experimental observations.

\paragraph*{Experimental setup.}

\begin{figure}
  \centering
  \includegraphics[width=\columnwidth]{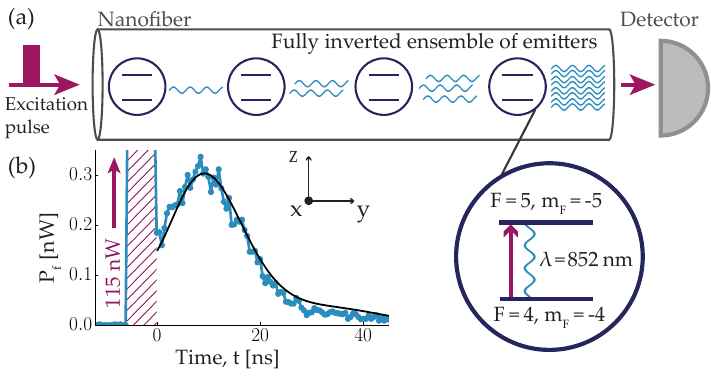}
\caption{(a) Schematic of the build-up of a superradiant burst in a cascaded system. Cesium atoms (blue circles) are trapped about 230~nm away from the nanofiber surface and are unidirectionally coupled to the evanescent field of the forward-propagating nanofiber-guided mode. This realizes a cascaded quantum system, where the dynamics of any atom can only influence downstream atoms. (b) We coherently invert an ensemble of about 1000 atoms using a forward-propagating nanofiber-guided optical pulse (purple arrow) and record the transmitted light. As the ensemble decays, we observe the emission of a superradiant burst into the forward-propagating mode. The solid black line represents the prediction of a cascaded interaction model, see main text.
}
\label{fig:setup}
\end{figure}
A schematic of the experimental setup is shown in Fig.~\ref{fig:setup}(a). We optically interface cesium atoms with the evanescent field surrounding an optical nanofiber, which is realized as the waist of a tapered optical fiber with a nominal diameter of 500~nm. Using nanofiber-guided light, we implement a two-color optical dipole trap that features two diametral arrays of trapping sites for the atoms, located about 230~nm from the fiber surface. We probabilistically load atoms into this trapping potential from a magneto-optical trap~\cite{vetsch2010optical}. Due to the collisional blockade effect, the maximum number of atoms per trapping site is limited to one~\cite{schlosser2002collisional, vetsch2012nanofiber}. We then apply a magnetic offset field of about 0.5~G in the $z$-direction. Using a fiber-guided laser field that is near-resonant with the $\ket{6S_{1/2}, F=4}\rightarrow\ket{6P_{3/2}, F=5}$ D2-transition, we apply side-selective degenerate Raman cooling (DRC) to the atoms in one of the arrays~\cite{meng2018near}, which are prepared in the outermost Zeeman state of the $F=4$ hyperfine ground state manifold, $\ket{g}=\ket{6S_{1/2}, F=4, m_F=-4}$. This process simultaneously removes the atoms from the opposite trap array.
After inferring the number of trapped atoms, $N$, via transmission spectroscopy~\cite{vetsch2010optical, supplemental}, we excite the atomic ensemble to the $\ket{e}=\ket{6P_{3/2}, F=5, m_F=-5}$ state using a 4~ns-long fiber-guided optical pulse that is much shorter than the excited state lifetime of about $30$~ns~\cite{steck}. The excitation laser field is locally $\sigma^-$-polarized~\cite{lekien2004field, mitsch2014quantum} and resonant with the $\ket{g}\rightarrow\ket{e}$ transition. We recently showed that this technique allows us to almost fully invert ensembles with up to 1000 atoms, with an excited state probability of about 80~$\%$~\cite{liedl2023collective}.  
Since the subsequently emitted light is $\sigma^-$-polarized, the probability that a single atom emits into the locally almost perfectly $\sigma^-$-polarized forward-propagating mode is $\beta_f\approx0.01$, about ten times larger than the probability for backward emission~\cite{mitsch2014quantum}.
We measure the power of the light that the atoms emit into the forward-propagating mode, $P_f$~\cite{supplemental}. 
To obtain sufficient counting statistics, we excite the atoms 400 times per sequence at a repetition rate of 5~kHz and average over several thousand sequences.
During this probing, we switch from our fiber-guided DRC as described above to DRC with a free-space laser beam that is near-resonant with the $\ket{6S_{1/2}, F=4}\rightarrow\ket{6P_{1/2}, F=4}$ D1-transition.
Since the corresponding scattering rate is much smaller than the collective decay rate of the atomic ensemble, we can continuously cool the atoms during the experimental sequence without disturbing their dynamics. 
Thanks to the continuous cooling, not more than 15~$\%$ of the atoms are lost during the probing sequence with a duration of 80~ms.

\paragraph*{Superradiant burst.}
In Fig.~\ref{fig:setup}(b), we show an example time trace of $P_f$ as blue dots. The $4$~ns excitation pulse is switched off at time $t=0$~ns.
Subsequently, the ensemble of about 1000 inverted atoms decays, and we observe an initial increase of $P_f$. The power reaches a maximum value of $P_f^\text{max}$ after a delay of $t_\text{D}\approx9$~ns and then decreases. This initial increase in emitted power is in stark contrast to the exponential decay of independent atoms and is a characteristic feature of a superradiant burst. We observe this burst  notwithstanding the fact that the atoms are chirally coupled to the nanofiber.
The synchronization of the atomic dipoles during their decay is enabled by the guided mode of the nanofiber, despite the large distance between atoms~\cite{tiranov2023collective}.
In addition, the unidirectional coupling makes the dynamics robust to the randomness of these distances.
Since the size of the Hilbert space grows exponentially with the number of atoms, a numerical solution of Eq.~\eqref{eq:master} is prohibitively costly for as many as $1000$ atoms.
Instead, we developed a cascaded interaction model whose computational complexity scales linearly with the number of atoms, the result of which is shown as a black dashed line in Fig.~\ref{fig:setup}(b). We describe this model, which agrees quantitatively with all data presented in this work, in detail further below.

\paragraph*{Scaling with the number of atoms.}

\begin{figure}
  \centering
  \includegraphics[width=\columnwidth]{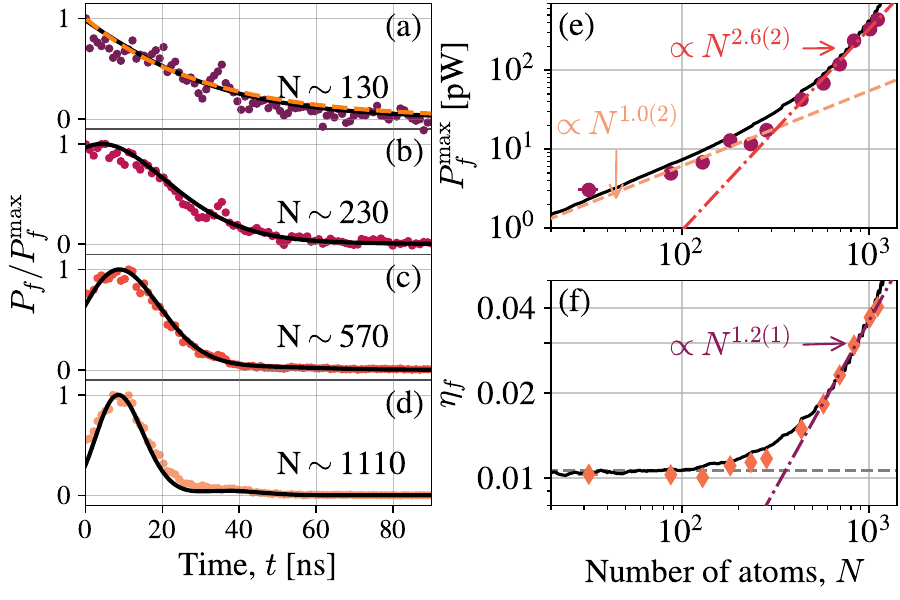}
\caption{(a)-(d) Power emitted by an inverted ensemble for different atom numbers, $N$. As we increase $N$, the dynamics changes from an exponential decay to a superradiant burst. (e) Scaling of the peak-emitted power, $P_\text{max}$, with $N$ (purple dots). We observe a linear scaling for small $N$ (orange dashed line), which becomes superlinear when $N$ exceeds a threshold of about 300 (red dash-dotted line). (f) Scaling of the fraction of absorbed energy emitted into the forward-propagating mode, $\eta_f$ (orange diamonds). Also here, we observe a threshold, below which $\eta_f$ is constant and above which it  increases with $N$. The corresponding model predictions are shown as solid black lines.}
\label{fig:burst_vs_N}
\end{figure}

The spontaneous synchronization of the atomic dipoles in our experiment is in competition with dephasing and decay into other modes~\cite{gross1982superradiance}. As a result, the superradiant burst only occurs when the number of atoms exceeds a threshold, beyond which spontaneous synchronization dominates.
To study this threshold, we measure the decay of fully inverted ensembles for varying atom number, $N$, see Fig.~\ref{fig:burst_vs_N}(a)-(d). 
For about $N=130$ in (a), we observe an exponential decay of $P_f$, identical to the case of independently emitting atoms.  
For $N=230$ in (b), we observe a plateau before $P_f$ eventually decreases, indicating that the system is at the onset of a superradiant burst. For $570$ atoms, a clear burst is apparent, which is even more pronounced for $1110$ atoms.
From these measurements, we extract the peak-emitted power, $P_f^\text{max}$, and plot it as a function of the atom number, $N$, in Fig.~\ref{fig:burst_vs_N}(e). Two regimes are clearly visible:
for small $N$, $P_f^\text{max}$ increases linearly, and a power-law fit reveals an exponent of 1.0(2).
In this regime, the peak power occurs directly after switching off the excitation laser. Since all atoms emit independently, the power is given by $P_f^\text{max}=P_f(t=0)=\Gamma_f E_\text{st}$, where $E_\text{st}\approx N\hbar\omega$ is the energy stored in the ensemble after the excitation pulse. This energy is radiated into the forward-propagating mode at a rate $\Gamma_f=\beta_f\Gamma$, where $\Gamma/(2\pi)\approx5.22$~MHz is the excited state decay rate.
As $N$ is increased further, the scaling changes, and the extracted exponent becomes 2.6(2). This threshold behavior and the superlinear scaling are indicative of a synchronization of the atomic dipoles, such that $P_f^\text{max}$ exceeds $\Gamma_f E_\text{st}$. Indeed, the change of scaling in $P_f^\text{max}$ is accompanied by the occurrence of burst dynamics, see Fig.~\ref{fig:burst_vs_N}(c) and (d).
In the textbook case of the emission of a Dicke-like superradiant burst by a strongly confined ensemble, the scaling is quadratic since each atom constructively adds to the macroscopic dipole moment of the ensemble, and the power is proportional to its square~\cite{dicke1954coherence}. 
There, the entire energy stored in the ensemble is emitted into a single spatial mode, regardless of whether the dipoles synchronize or not.
However, in our experiment, a single atom emits only about 1\,$\%$ of its radiation into the (forward-propagating) guided mode, and about 99\,$\%$ of the light is scattered into the free-space modes. During the collective decay of the atoms, a phase pattern is formed across the atomic dipoles, which leads to constructive interference for emission into the forward-propagating mode. 
Therefore, the atoms emit into a narrower solid angle as $N$ increases. Accordingly, a larger fraction of the radiated power is collected by the nanofiber-guided mode. 
This explains the faster-than-quadratic scaling beyond the threshold observed in our experiment. 
In contrast to the weakly excited regime, where a similar phase pattern can be imprinted on the ensemble by an external laser field~\cite{solano2017super, corzo2019waveguide,okaba2019superradiance, pennetta2022observation}, here, this pattern forms spontaneously, as we show further below.
We further analyze the collective enhancement of forward scattering by measuring the fraction of the stored energy that is emitted into the forward-propagating mode, $\eta_f$~\cite{liedl2023collective}. 
We plot $\eta_f$ as a function of $N$ in Fig.~\ref{fig:burst_vs_N}(f) and find a similar threshold behavior as for the peak power, $P_f^\text{max}$. For small $N$, $\eta_f$ stays at a constant value of about $0.01\approx \beta_f$, as expected for independently emitting atoms, see dashed gray line in Fig.~\ref{fig:burst_vs_N}(f). For larger atom numbers, $\eta_f$ increases as $N^{1.2(1)}$, i.e., forward scattering is indeed collectively enhanced compared to independent decay. 

\paragraph*{Dependence on the initial state.}

\begin{figure}
  \centering
  \includegraphics[width=\columnwidth]{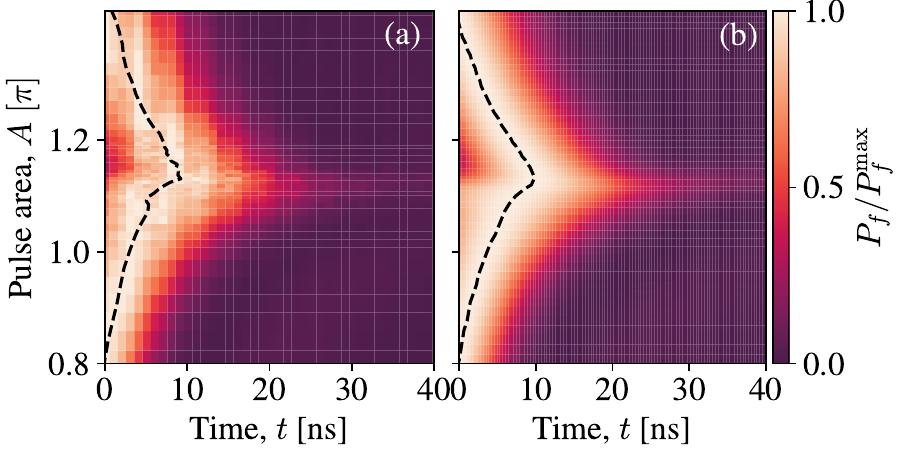}
\caption{Dynamics of the power emitted into the nanofiber-guided mode, $P_f$, for different pulse areas seen by the first atom, $A$. Each time trace is normalized to its peak value, $P_f^\text{max}$. The time at which the maximum is reached, $t_\text{D}$, is indicated by the black dashed lines. (a) shows the measured data, while (b) shows the corresponding model predictions.}
\label{fig:burst_vs_power}
\end{figure}

We now explore the dynamics of the superradiant burst for different initial states of the atomic ensemble consisting of about $1000$ atoms. By exciting the atoms with short, resonant optical pulses of varying power and, correspondingly, different pulse areas, $A$, we prepare each atom to good approximation in a coherent superposition of ground and excited state~\cite{liedl2023collective},
\begin{equation}
\ket{\psi}=\cos\left(\frac{A}{2}\right)\ket{g}-\imu\sin\left(\frac{A}{2}\right)\ket{e},
\label{eq:psi0}
\end{equation}
where we have chosen a frame that rotates with the forward-propagating laser mode. 
The time traces of the emitted power, $P_f$, are shown as a function of $A$ in Fig.~\ref{fig:burst_vs_power}(a), where each time trace is normalized to its peak value, $P_f^\text{max}$. 
We extract the time at which the maximal power is emitted into the nanofiber, $t_\text{D}$ and display it as the black dashed line. 
For small pulse areas ($A<0.8\pi$), the power decreases monotonously as the ensemble decays, and $t_\text{D}=0$. 
However, for $A$ between about $0.8\pi$ and $1.4\pi$, the ensemble is substantially inverted and a superradiant burst is apparent, such that $t_\text{D}>0$. The closer the ensemble is to full inversion, the larger the delay $t_\text{D}$, with a maximal value of $t_\text{D}\approx 9$~ns. 
When $A$ is further increased beyond $1.4\pi$, the ensemble is coherently de-excited and the burst gradually vanishes.
From Eq.~\eqref{eq:psi0}, one would expect that the maximal inversion and thus the largest delay appears at $A=\pi$, and that the dynamics is symmetric below and above $A=\pi$. From our data, however, we find that the dynamics are slightly asymmetric about the point of maximal inversion, which appears at $A=1.13\pi$.
This is a consequence of the finite absorption of the excitation pulse along the ensemble and the cascaded interaction~\cite{supplemental, liedl2023collective}. 
In Fig.~\ref{fig:burst_vs_power}(b), we show the corresponding model prediction and find quantitative agreement, confirming that our cascaded interaction model captures the essential physical mechanism which determines the system dynamics. Further details on how imperfections influence the burst dynamics can be found in the Supplemental Material~\cite{supplemental}.

\paragraph*{First-order coherence properties.}

\begin{figure}
  \centering
  \includegraphics[width=\columnwidth]{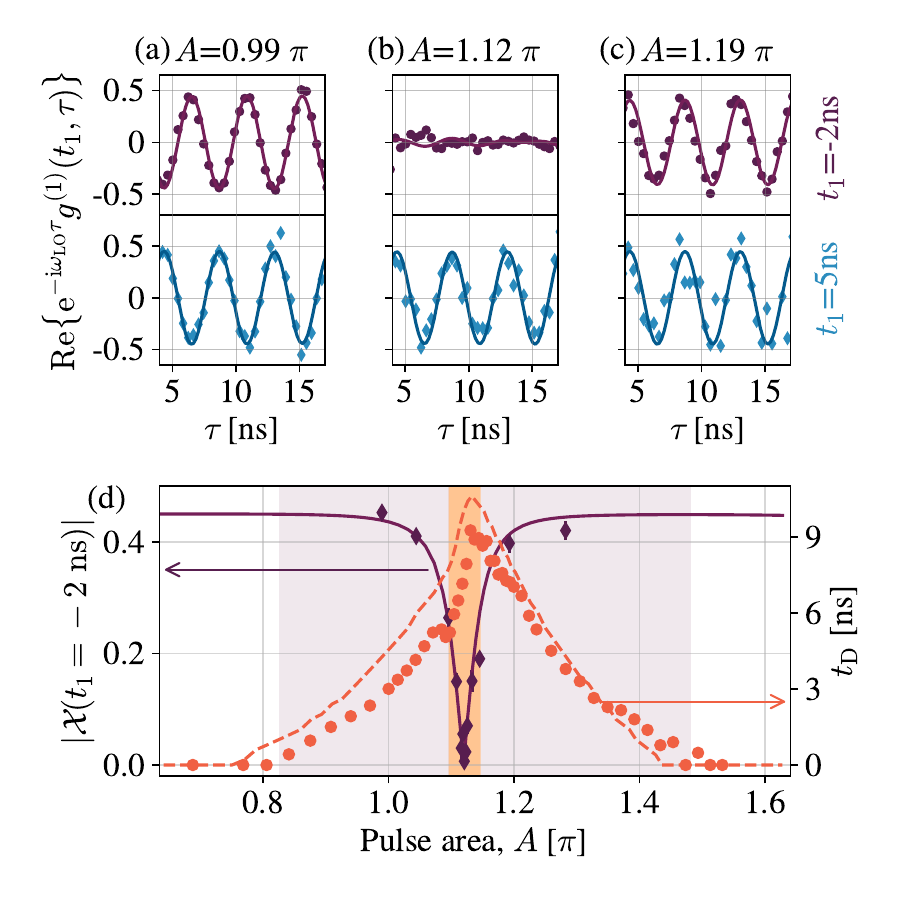}
\caption{Heterodyne analysis of the first-order coherence, $g^{(1)}(t_1,\tau)$, of the forward-emitted light, modulated with the local oscillator detuning of $\omega_\text{LO} = 2\pi\times230$~MHz as a function of the pulse area $A$.
 In the top of panels (a) to (c), $t_1=-2$~ns lies within the excitation pulse. The $g^{(1)}$ function is thus given by the cross-correlation of the excitation laser and the burst light.
 In (b), when maximal inversion is reached, the oscillations disappear, indicating that there is no fixed phase relationship between fluorescence and excitation laser.
In the bottom row panels of (a) to (c), $t_1=5$~ns lies within the burst. The interference fringes indicate that the superradiant emission is first-order coherent for  all considered values of $A$.
 (d) Magnitude of the coherence of the emitted light with respect to the laser, $|\mathcal{X}|$, for different $A$ (purple diamonds on left axis), and time delay of the superradiant burst, $t_\text{D}$ (orange dots on right axis).
}
\label{fig:heterodyne}
\end{figure}

To understand the role of coherence of the superradiant burst in our system, 
we measure the complex-valued normalized first-order coherence function of the forward-propagating light, $g^{(1)}(t_1,\tau)$.
Since we analyze a non-stationary field, $g^{(1)}(t_1,\tau)$ depends explicitly on two times, $t_1$ and $t_1+\tau$~\cite{loudon2000quantum} (the excitation pulse is switched off at $t_1=0$, cf.~Fig.~\ref{fig:setup}).
We superpose the light emitted into the forward-propagating mode with a local oscillator (LO) that is derived from the excitation laser and detuned by $\omega_\text{LO}=2\pi\times 230$~MHz. 
The intensity-intensity correlations of the resulting heterodyne signal contain a beating pattern, whose visibility is a measure of $g^{(1)}(t_1,\tau)$. Specifically, the normalized beating pattern is given by $\re{\left\{\expu^{-\imu \omega_\text{LO}\tau} g^{(1)}(t_1, \tau) \right\}}$, where $\re{\left\{\ldots\right\}}$ denotes the real part~\cite{supplemental}.  
In Fig.~\ref{fig:heterodyne} (a) to (c), we plot the measured beating patterns for three different pulse areas, $A$, and for two different times $t_1$.
Throughout the considered parameter space, we observe that the beating pattern is well described by a function of the form $\mathcal{X}(t_1)\cos(\omega_\text{LO}\tau)$, i.e., $g^{(1)}(t_1,\tau)$ is real-valued. The amplitude $\mathcal{X}(t_1)$ thus quantifies the time-averaged first-order coherence.
While full first-order coherence would in principle lead to $|\mathcal{X}(t_1)|=1$, a polarization mismatch between the LO and the signal field reduces the visibility  in our experiment, so that a fully coherent signal field only leads to $|\mathcal{X}(t_1)|\approx0.5$, see Supplemental Material for details~\cite{supplemental}.

In the top of panels (a) to (c), we set $t_1=-2$~ns (i.e., when the excitation pulse is on) and $\tau>2$~ns (i.e., during the superradiant burst). 
Consequently, the interference signal corresponds to a cross-correlation between the excitation laser light and the superradiant burst emitted by the atoms.
In panel (a), the pulse area is slightly below the point of maximal inversion, and we observe a beating pattern with  near maximum visibility, $\mathcal{X}(t_1)\approx -0.5$, indicating that the emitted light is coherent with respect to the excitation laser field. Here,  $\mathcal{X}(t_1)$ is negative, because the burst light emitted by the atoms is phase shifted by $\pi$ with respect to the laser, as expected.
In panel (c), the ensemble is coherently driven beyond full inversion, and the atoms radiate in phase with the laser, indicated by $\mathcal{X}(t_1)\approx +0.5$.
Importantly, however, for maximal inversion as shown in (b), we do not observe any interference fringes, $\mathcal{X}(t_1)\approx 0.0$, i.e., the emitted light features no fixed phase relationship with respect to the excitation laser field.
In  Fig.~\ref{fig:heterodyne}(d), we plot $|\mathcal{X}(t_1=-2~{\rm ns})|$ as a function of the pulse area $A$ (purple diamonds).
In addition, we show the delay time, $t_\text{D}$, at which the peak power $P_f^\text{max}$ is emitted (orange dots). This is the same data shown as a black dashed line in Fig.~\ref{fig:burst_vs_power}(a). We observe that the minimum of $|\mathcal{X}|$ appears when the delay $t_\text{D}$ is longest, i.e., when the ensemble is maximally inverted. Interestingly, the coherence with respect to the laser features a dip which is much narrower than the region in which $t_\text{D}$ is larger than zero. This allows us to identify two regimes of superradiant burst dynamics in our experiment.
In the first regime close to full inversion, indicated by the light-red shaded area, the superradiant burst is incoherent with respect to the excitation laser and is therefore triggered by vacuum fluctuations.
In this case, one sometimes speaks of superfluorescence~\cite{bonifacio1975cooperative}. In a second regime comprising a much broader range of pulse areas indicated by the light-purple shaded regions, the burst is coherent with the excitation laser. Here, the phase is imprinted onto the ensemble by the excitation laser field. 

In the lower row of panels (a) to (c), we set $t_1=5$~ns (i.e., after the excitation pulse has been switched off), thereby measuring the first-order coherence of the superradiant burst. 
We observe that $\mathcal{X}(t_1)\approx 0.5$ for all three excitation pulse areas $A$, including for $A=1.12\pi$ in panel (b), where the atomic ensemble is maximally inverted, has no total dipole moment, and is thus incoherent with respect to any external reference. 
This indicates that even the vacuum-induced superradiant burst is predominantly emitted into a single temporal mode~\cite{fabre2020modes}.
Such a behavior has recently been predicted for Dicke superradiance of ensembles of two-level systems that are coupled symmetrically~\cite{perarnau2020multimode, lemberger2021radiation}, but has, to our knowledge, not been confirmed experimentally so far.\\
Let us now discuss the observation of an interference pattern in the heterodyne measurement of the vacuum-induced burst in the context of the theory of heterodyne detection of a quantum mechanical, single field mode $\hat E$. In a Heisenberg picture, where quantum operators depend on time,  the autocorrelation function of the heterodyne photocurrent is a measure for the correlator $G^{(1)}(t_1,\tau) = \braket{\dg{\hat E}(t_1+\tau) \hat E(t_1)}$~\cite{carmichael1987spectrum}, as defined by Glauber~\cite{glauber1963quantum}.
In an equivalent Schr{\"o}dinger picture, where operators are time-independent, however, this interference seems to be a consequence of a non-vanishing dipole moment appearing in the conditional atomic quantum state during the heterodyne measurement~\cite{wiseman1993quantum, wiseman2009quantum, bolund2014stochastic}. 
While a quantitative modelling of our experiment along these lines is beyond the scope of the present work, we do present a theoretical study of a fully inverted ensemble of independently emitting atoms in the Supplemental Material~\cite{supplemental}. Interestingly, also in this situation, the emitted field is predicted to feature full coherence, similar to our observations.

\paragraph*{Theoretical model.}\label{sec:model}
Let us now turn to the theoretical description of our system.
The many-body master equation for the density operator $\hat\rho$ of $N$ atoms, which are coupled to a unidirectional waveguide is given by~\cite{lekien2008cooperative, stannigel2012driven, pichler2015quantum, mahmoodian2020dynamics, cardenas2022many}
\begin{equation}\label{eq:master}
\frac{\diff }{\diff t} \hat  \rho = -\frac{\imu}{\hbar} [\hat H_\text{casc}, \hat \rho] +  \Gamma_f \mathcal{L}_\text{coll}[\hat \rho] + \Gamma_0 \mathcal{L}_0[\hat \rho]
\end{equation}
with the cascaded interaction Hamiltonian,
\begin{equation}\label{eq:hamiltonian}
\hat H_\text{casc} = -\imu\frac{ \hbar\Gamma_f}{2} \sum_{ k<l} \dg{\hat \sigma}_l \hat \sigma_k + H.c.
\end{equation}
Here, 
$\hat \sigma_k=\ket{g}_k\bra{e}_k$ is the spin lowering operator of the $k$th atom in a frame which co-propagates with the guided mode with $k=1,\dots,N$. The indices are increasing in the direction of the propagating mode. Importantly, this Hamiltonian only allows information to propagate in one direction along the waveguide.
The collective decay into the waveguide with rate $\Gamma_f=\beta_f\Gamma$ is described by Lindblad superoperator $\mathcal{L}_\text{coll}[\hat \rho] $, while independent decay into free space with rate $\Gamma_0=(1-\beta)\Gamma$ is described by $\mathcal{L}_0[\hat \rho]$,
\begin{subequations}
\begin{align}
	\mathcal{L}_\text{coll}[\hat\rho] &= \frac{1}{2}\sum_{k , l} \left(2 \hat \sigma_l \hat\rho \dg{\hat \sigma}_k - \dg{\hat \sigma}_k \hat \sigma_l  \hat\rho -   \hat\rho \dg{\hat \sigma}_k \hat \sigma_l   \right), \\
	\mathcal{L}_0[\hat\rho] &= \frac{1}{2}\sum_{k} \left(2 \hat \sigma_k \hat\rho \dg{\hat \sigma}_k - \dg{\hat \sigma}_k \hat \sigma_k  \hat\rho -   \hat\rho \dg{\hat \sigma}_k \hat \sigma_k   \right).
\end{align}
\end{subequations}
We note that while the many-body master equation, Eq.~\eqref{eq:master}, is useful for understanding the physical properties of the system, its solution is inaccessible for as many as $N=1000$ coupled atoms, since the size of the density matrix $\hat \rho$ scales exponentially with $N$.
In order to approximate the solution numerically, let us note that the dynamics of each individual atom in the ensemble is described by the quantum Langevin equation~\cite{gardiner1985input}
\begin{equation}\label{eq:Langevin}
\frac{\diff}{\diff t} \hat \sigma_k = -\frac{\Gamma}{2} \hat \sigma_k -\imu \sqrt{\Gamma_f}\left(1-2\dg{\hat\sigma}_k\hat\sigma_k\right)\hat a_k(t),
\end{equation}
where $\hat a_k(t)$ is the field operator of the waveguided field before the $k$th atom. 
Because of the cascaded interaction, the output field of this mode serves as the next atom's input, $\hat a_{k+1}(t)$, and is given by the input-output equation~\cite{gardiner1985input}
\begin{equation}
\hat a_{k+1}(t) = \hat a_k(t) - \imu  \sqrt{\Gamma_f}\hat \sigma_k.
\end{equation}
For our cascaded quantum system, these equations are equivalent to the master equation~\eqref{eq:master}~\cite{gardiner1993driving, carmichael1993quantum}.
In Ref.~\cite{liedl2023collective}, we approximated the photonic state between the atoms as coherent states, i.e., we replaced $\hat a_k(t)$ by its amplitude $\alpha_k(t)=\braket{\hat a_k(t)}$. This reduces the Langevin equation of the $k$th atom to the (single-atom) optical Bloch equations and allows one to solve the mean-field atomic dynamics of the whole ensemble with linear computational complexity.
However, it is well known that the output field $\hat a_{k+1}$ is in general no longer in a coherent state. In particular, close to full inversion there is no coherence left.
Therefore, the total output flux, $P_{k+1} (t) = \braket{\dg{\hat a}_{k+1}\hat a_{k+1}}(t)$, is always larger than the coherent part, $P^c_{k+1} (t) = |\braket{\hat a_{k+1}}(t)|^2$.
The difference $P^\text{inc}_{k+1} (t)= P_{k+1} (t) - P^c_{k+1}(t)$ is commonly referred to as the ``incoherent'' part of the emitted light, which is due to spontaneous emission.
In order to describe this incoherent driving of the next atom, we model the input field $\hat a_{k}(t)$ as the superposition of a coherent field and a ``randomly-phased laser field''~\cite{mandelwolf1995optical}. The density operator of such a field is given by
\begin{equation}
\hat \rho_L(t) = \frac{1}{2\pi}\int_0^{2\pi}\diff\phi ~ \ket{\alpha(\phi,t)}\bra{\alpha(\phi,t)},
\end{equation}
which represents a mixture of coherent states with amplitudes $\alpha(\phi,t) = \sqrt{P_k^c(t)} + \expu^{\imu\phi} \sqrt{P_k^\text{inc}(t)}$, homogeneously sampled over the phase $\phi$.
This ``mixed coherent state'' has the necessary property that its total power $\braket{\dg{\hat a}_k \hat a_k }(t) = P_k^c(t) + P_k^\text{inc}(t)$ is larger than its coherent part $|\braket{\hat a_k}(t)|^2=P_k^c(t)$. Moreover, we can still straightforwardly solve the atomic dynamics of the $k^\text{th}$ atom given this input state by solving the optical Bloch equation for each phase $\phi$. We then numerically compute the output's coherent and incoherent parts $P_{k+1}^c(t)$ and $P_{k+1}^\text{inc}(t)$, respectively, by integrating over $\phi$.
We apply this method iteratively to each atom and finally obtain the output power $P_f (t) = P_{N+1}(t)$.
The computational complexity of this method is thus linear in $N$.

In addition, we average the simulated time traces over a truncated Gaussian distribution of $\beta_f$-values to account for thermal fluctuations of the atomic position relative to the nanofiber surface and fit the corresponding mean and standard deviation to the experimental data, yielding $\bar\beta_f=0.0112$ and $\sigma
_\beta=0.0065$, respectively~\cite{liedl2023collective, supplemental}. Notably, we only need these two free parameters to obtain quantitative agreement with the data throughout the whole parameter space studied in this work.  
We numerically confirmed that the finite duration of the excitation pulse and the inhomogeneous spread of coupling strengths does not qualitatively alter the observed dynamics~\cite{supplemental}. The quantitative  agreement of our model with the data attests that our system can indeed be modeled as a cascaded quantum system and that our mixed coherent state approximation is justified.
Other methods that account for incoherent dynamics beyond the mean field typically rely on higher order cumulant expansions~\cite{kusmierek2023higherorder}. To our knowledge, while these models are solvable in polynomial time, the solution for more than a few hundred atoms remains elusive.

\paragraph*{Conclusions and outlook.}

In conclusion, we have experimentally observed superradiant burst dynamics of an ensemble of atoms that is unidirectionally coupled to a guided mode of an optical nanofiber. Our results demonstrate that superradiance prevails in a cascaded quantum system, despite the reduced symmetry of the light--matter coupling.
The scaling of the peak power of the burst emission with the number of atoms is observed to be faster than in the Dicke case, which we could explain in an intuitive way. Lastly, we presented a direct measurement of the coherence of the burst emission and its dependence on the initially prepared atomic state, experimentally demonstrating a textbook prediction. Notably, this allowed us to show that the superradiant burst is predominantly emitted into a single temporal mode, thereby demonstrating an important prerequisite for generating multi-photon quantum states with high fidelity, which may, e.g., turn out useful in quantum metrology~\cite{gonzalez2015deterministic, paulisch2019quantum, perarnau2020multimode, tziperman2023spontaneous}.
All data shown in this work are in quantitative agreement with a model prediction, which is made possible by the fact that we deal with a cascaded quantum system, resulting in a drastic reduction of the computational complexity compared to the exponential scaling of the many-body master equation.

Future research directions include the experimental investigation of the second-order coherence of the burst emission~\cite{jahnke2016giant}. It would also be interesting to study the long-term dynamics of the decay. Here, sub-radiant features related to a high degree of correlations between the atomic dipoles are expected. For continuous driving, a super-radiant phase transition has recently been observed with an atomic ensemble emitting into free space~\cite{ferioli2023non}. When chirally coupled atoms are driven continuously, steady-state many-body entanglement between the emitters is predicted to occur~\cite{stannigel2012driven,ramos2014quantum}. The search for and experimental investigation of robust entanglement signatures in systems like ours therefore constitute another promising research avenue. Finally, placing the atoms in a nanofiber-integrated optical resonator would yield a versatile testbed for studying the physical mechanism underlying a superradiant laser~\cite{meiser2009prospects, bohnet2012steady}.

\begin{acknowledgments}
We thank A.~Asenjo García, A.~Browaeys, S.~Cardenas López, J.~I.~Cirac, I.~Ferrier-Barbut, K.~Hammerer, K.~Kusmierek, S.~Masson, K. M{\o}lmer, L.~A.~Orozco, J.~Volz, and L.~P.~Yatsenko for stimulating discussions and helpful comments. We acknowledge funding by the Alexander von Humboldt Foundation in the framework of the Alexander von Humboldt Professorship endowed by the Federal Ministry of Education and Research, 
as well as funding by the European Commission under the project DAALI (No. 899275).
\end{acknowledgments}

\bibliography{bibliography}
\clearpage

\renewcommand{\theequation}{S\arabic{equation}}
\renewcommand{\thefigure}{S\arabic{figure}}
\setcounter{equation}{0}
\setcounter{figure}{0}

\onecolumngrid
\begin{center}
\large{\bf Supplemental Material:\\Observation of superradiant bursts in a cascaded quantum system}\\
\vspace{0.6cm}
\end{center}
\twocolumngrid

\subsection{Loading of the nanofiber-based atom trap}
We load cesium atoms into a magneto-optical trap (MOT) and perform optical molasses cooling to transfer the atoms into the nanofiber-based trapping potential, which features two diametral arrays of trapping sites~\cite{vetsch2010optical}. The trapping potential is created by running-wave blue-detuned nanofiber-guided field (wavelength 760 nm, power 20.5 mW) and a standing-wave red-detuned field (wavelength 1064 nm, total power 2.4 mW). The loading is probabilistic and, due to the collisional blockade effect, each potential minimum is occupied by at most one atom~\cite{schlosser2002collisional}. Then, we apply a homogenous magnetic offset field of about 0.5~G along 
$+z$ (see Fig.~1 in the main text) and further cool the atoms on one side of the nanofiber by degenerate Raman cooling (DRC) using nanofiber-guided light that is near-resonant with the $\ket{6S_{1/2}, F = 4}\to \ket{6P_{3/2}, F = 5}$ D2 transition~\cite{meng2018near}. At the same time, the atoms on the other side of the nanofiber are subject to degenerate Raman heating and are expelled from the trap, such that we are left with a one-dimensional array of atoms on only one side of the nanofiber.
After these preparatory steps, we switch off both the MOT and DRC lasers fields such that during the subsequent measurements, there is no resonant light at a wavelength of $852$~nm, except for the excitation laser. As described in the main text, we then use an additional free-space laser resonant with the D1 transition at a wavelength of $894$~nm in order to cool the trapped atoms by DRC.
The number of trapped atoms, $N$, can be tuned by changing the MOT loading time, which ranges from 80~ms to 7~s.

\subsection{Detection setup}
\begin{figure}
  \centering
	\includegraphics[width=0.8\columnwidth]{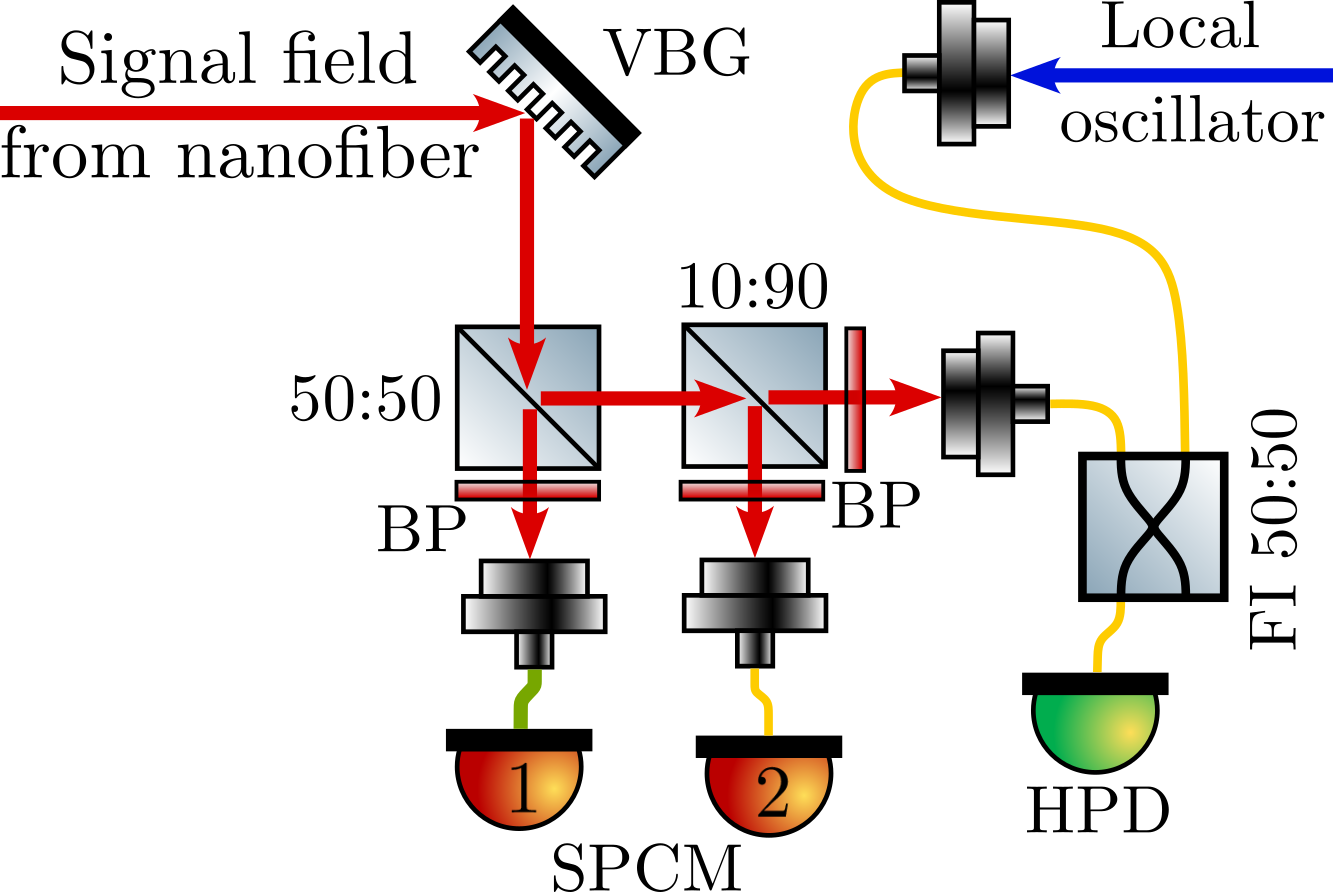}
	\caption{Schematic of the detection setup. The signal field is spectrally filtered around a wavelength of 852~nm using a volume Bragg grating (VBG) and a bandpass filter (BP) in front of each fiber-coupled detector. The yellow and green lines depict single-mode and multi-mode fibers, respectively. We measure the power of the signal field using two single photon counting modules (SPCM) and a hybrid photodetector (HPD). The light incident on the HPD can be overlapped with a local oscillator using a fiber-integrated 50:50 beamsplitter (FI 50:50). The frequency of the local oscillator field is shifted by 230~MHz with respect to the excitation laser field.}
\label{fig:Detection_setup}
\end{figure}

Figure~\ref{fig:Detection_setup} schematically depicts the detection setup that we use to measure the power of the light exiting the waveguide in the forward direction. We first spectrally filter the signal field at a wavelength of 852~nm from the trapping light fields and other background light using a volume Bragg grating. The light is first split up by a 50:50 beam splitter and then further separated by a 10:90 $(R:T)$ beam splitter. To further suppress the background, each beam then passes a bandpass filter that is centered around 852~nm. We detect the light at the different output ports of the beam splitters using two fiber-coupled single photon counting modules (SPCM~1 and SPCM~2) and a fiber-coupled hybrid photodetector (HPD). We use SPCM~1 to measure the probe transmission spectrum, from which we infer the optical depth of the atomic ensemble and the number of trapped atoms. 
For the experiments shown in the manuscript, we expose the atomic ensemble to intense optical pulses, which have a power that is several orders of magnitude larger than the power emitted by the ensemble during its decay. Since our SPCMs have a dead time of about 25~ns, we have to substantially attenuate the power incident on the SPCMs in order to detect the fluorescence that follows the intense excitation pulse. In contrast, the HPD features a much shorter dead time of only about 2~ns. We can therefore saturate the HPD during the excitation pulse and still detect the fluorescence signal.  The signal of the HPD is highly saturated during the excitation pulse. Therefore, we use the signal from SPCM~2 to extract the fraction of absorbed energy that is emitted into the forward direction by the atoms, $\eta_f$. To ensure that the excitation pulse does not saturate the signal from SPCM~2, we attenuate the light incident on SPCM~2 using neutral density filters.\\
For the measurements presented in Sec.~V of the main manuscript, we superimpose the signal incident on the HPD with a local oscillator light field using a fiber-integrated 50:50 beam splitter. We derive the local oscillator field from the excitation laser field and shift its frequency by 230~MHz. While performing the heterodyne measurements, we also record the bare signal field (without local oscillator) using SPCM~2. We use the resulting time traces to normalize the first-order coherence function that we extract from the heterodyne signal, see below.

\subsection{Extraction of the first-order coherence function from the heterodyne signal}
The first- and second-order correlation functions of a light field $\hat a(t)$ with power $P(t)=\langle \hat a^\dagger(t)\hat a(t)\rangle$ are defined as~\cite{loudon2000quantum}
\begin{equation}
\begin{split}
    G^{(1)}(t,\tau) &=\langle \hat a^\dagger(t)\hat a(t+\tau)\rangle, \\
    g^{(1)}(t,\tau) &= \frac{G^{(1)}(t,\tau)}{\sqrt{ P(t) P(t+\tau)}}, \\
    G^{(2)}(t,\tau) &=\langle\hat  a^\dagger(t)\hat a^\dagger(t+\tau)\hat a(t+\tau)\hat a(t)\rangle, \\
    g^{(2)}(t,\tau) &= \frac{G^{(2)}(t,\tau)}{ P(t)P(t+\tau)}. 
\end{split}
\label{eq:definition_gs}
\end{equation}
Here, $g^{(1)}$ and $g^{(2)}$ denote the normalized first-order and second-order correlation functions, respectively. 
Note that for a non-stationary case, such as we analyze, these correlators explicitly depend on two time instances, $t$ and $t+\tau$.
Let us assume a light field $\hat a(t)$ with power $P(t)$ that describes the nanofiber-guided mode and superpose a classical, continuous-wave local oscillator field $a_\text{LO}$ with power $P_\text{LO}$, with a random relative phase, $\theta_\text{LO}$, and relative frequency, $\omega_\text{LO}$,
\begin{align}
    a_\text{LO}(t) &= \sqrt{P_\text{LO}}e^{i(\omega_\text{LO}t+\theta_\text{LO})}.
\end{align}
The field incident on the detector and its power are then given by
\begin{align}
    a_D(t) &= \sqrt{P_\text{LO}}e^{i(\omega_\text{LO}t+\theta_\text{LO})} + a(t), \\
    P_D(t) &=P_\text{LO}+P(t).
    \label{eq:P_D}
\end{align}
In our experiment, we measure the normalized second-order correlation function of this heterodyne signal. We average over the relative phase between local oscillator and the signal field, $\theta_\text{LO}$, since the latter is not stabilized and drifts randomly on the time scale of the repetition period of our measurements. Due to these random drifts, there is no interference term in Eq.~(\ref{eq:P_D}) since it averages to zero. Using the definitions in Eqs.~(\ref{eq:definition_gs}), we obtain the normalized second-order autocorrelation function of the heterodyne signal,
\begin{equation}
\begin{split}
g^{(2)}_D(t, \tau) = &1+ \frac{2 P_\text{LO}\sqrt{P(t)P(t+\tau)}}{P_D(t)P_D(t+\tau)} \re{ \left\{\expu^{-\imu \omega_\text{LO}\tau} g^{(1)}(t, \tau) \right\} }\\
&+\frac{P(t)P(t+\tau)}{P_D(t)P_D(t+\tau)}[g^{(2)}(t,\tau)-1].
\end{split}
\end{equation}
Here, $g^{(1)}$ and $g^{(2)}$ refer to the first- and second-order coherence functions of the signal field $\hat a(t)$, as defined above.
Since, for the experimental parameters in our experiment, \mbox{$P_\text{LO}\gg\sqrt{P(t)P(t+\tau)}$}, we neglect the last term that includes the contribution of $g^{(2)}(t,\tau)$ such that we are left with

\begin{align}
    g^{(2)}_D(t, \tau) = 1+ V_\text{max}(t, \tau) \re{ \left\{\expu^{-\imu \omega_\text{LO}\tau} g^{(1)}(t, \tau) \right\} }.
\end{align}
Here, we have introduced the maximum visibility of $V_\text{max}(t, \tau)$. 
\begin{align}
    V_\text{max}(t, \tau) = \frac{2 P_\text{LO}\sqrt{P(t)P(t+\tau)}}{P_D(t)P_D(t+\tau)}.
\end{align}
The second-order correlation function of our heterodyne signal is thus given by the first-order correlation function of the signal field $a(t)$, oscillating at $\omega_\text{LO}$. 
We extract both $g_D^{(2)}(t,\tau)$ and $V_\text{max}(t, \tau)$ from our measurements and infer the following quantity from our data: 
\begin{align}
    \re{ \left\{\expu^{-\imu \omega_\text{LO}\tau} g^{(1)}(t, \tau) \right\} } = \frac{g^{(2)}_D(t, \tau)-1}{V_{max}(t, \tau)},
\end{align}
which we show in Fig.~4 of the manuscript. 
Note that in this work we drive the atomic ensemble on resonance such that $g^{(1)}$ is purely real. As discussed in the main manuscript, it can be both negative (for excitation pulse areas below $\pi$) and positive (for excitation pulse areas above $\pi$).
Due to an imperfect overlap between the polarizations of the signal field and the local oscillator field, the maximal visibility in the experiment is further reduced. We experimentally checked that the oscillations shown in Fig.~4 of the main manuscript indeed reach maximal contrast if a polarization filter is introduced in front of the detector.

\subsection{First-order coherence of independently decaying atoms}
Consider a single two-level system and lowering operator $\hat\sigma = \ket{g}\bra{e}$, which is initially excited at time $t=0$, and which decays through spontaneous emission with rate $\gamma$. Using standard techniques, we find the following time-dependent correlators~\cite{carmichael1999statistical},
\begin{subequations}
\begin{align}
\braket{\hat\sigma(t)} &= 0,\\
\braket{\dg{\hat \sigma}(t)\hat\sigma(t)} &= \expu^{-\gamma t},\\
\label{eq:G1_single}
\braket{\dg{\hat  \sigma}(t)\hat \sigma(t+\tau)} &= \expu^{-\gamma t - \gamma \tau /2 }.
\end{align}
\end{subequations}
Here, we are in a rotating frame at the transition frequency of the atom.
If we now consider the emitted field $\hat a= r \hat\sigma$ with some coupling constant $r$, we find the normalized, first-order coherence function of $\hat a(t)$ as
\begin{equation}
g^{(1)}(t,\tau) = \frac{\braket{\dg{\hat \sigma}(t)\hat\sigma(t+\tau)} }{\sqrt{\braket{\dg {\hat \sigma} \hat \sigma}(t) \braket{\dg {\hat\sigma}\hat\sigma}(t+\tau) }} = 1.
\end{equation}
The field emitted by a single, excited atom is thus first-order coherent, i.e., it is emitted into a single temporal mode~\cite{fabre2020modes}. 

Next, we consider $N$ atoms which are all prepared in the excited state. Let us couple the scattered light of the atoms into a single optical mode, $\hat a$, which is given by 
\begin{equation}
\hat a(t) = \sum_{n=1}^N r_n \hat\sigma_n(t).
\end{equation}
Here, $r_n$ are the individual coupling constants of the atoms to the detection mode and $\hat\sigma_n$ is the lowering operator of the $n$th atom. We assume independent dynamics of the atoms, that is the emitted field of any atom can not influence the dynamics of any other atom. Because of the independence, we have for $m\neq n$: $\braket{\dg{\hat \sigma}_m(t) \hat\sigma_n(t+\tau)} = \braket{\dg {\hat \sigma}_m(t)} \braket{ \hat \sigma_n(t+\tau)} = 0$. With this, we find the first-order coherence function of $\hat a(t)$ as
\begin{subequations}
\begin{align}
\braket{\dg {\hat a}(t)\hat a(t+\tau)} &= \sum_{n=1}^N |r_n|^2 \braket{\dg{\hat  \sigma}_n(t)\hat\sigma_n(t+\tau)}.
\end{align}
\end{subequations}
Since all atoms are excited at the same time, the correlators $\braket{\dg{\hat \sigma}_n(t)\hat\sigma_n(t+\tau)}$ are identical to Eq.~\eqref{eq:G1_single} and independent of $n$. From this it follows directly, that also in the case of many, independent atoms, $g^{(1)}(t,\tau) = 1$.

\subsection{Fluctuations of the coupling strength due to temperature}
\begin{figure}
  \centering
	\includegraphics[width=0.8\columnwidth]{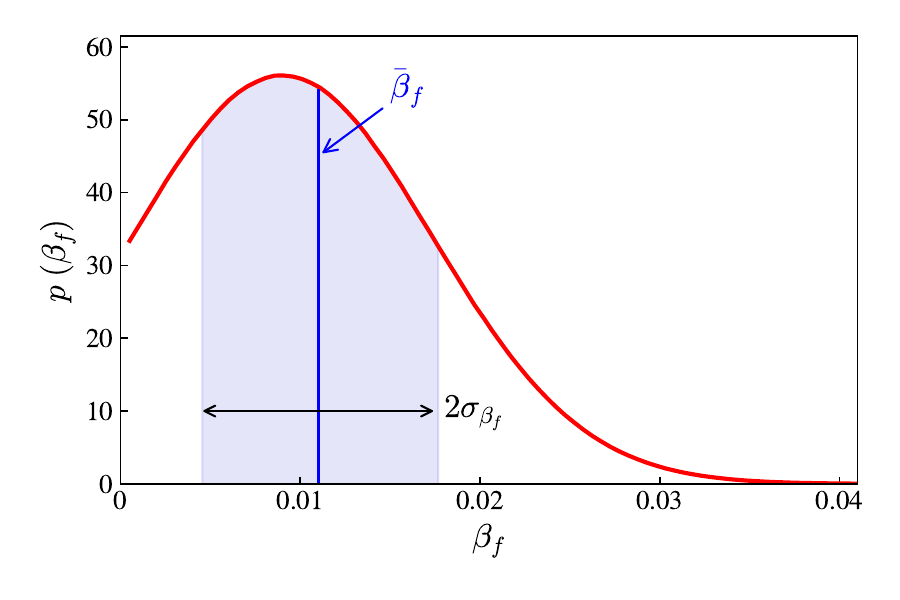}
	\caption{Probability distribution used to model temperature-induced fluctuations of the atom-waveguide coupling strength, $\beta_f$. The fitted mean value, $\bar\beta_f=0.0112$, and the standard deviation, $\sigma_{\beta_f}=0.0065$, of the truncated Gaussian distribution are shown as the blue vertical line and black double arrow, respectively.}
\label{fig:dist_beta}

\end{figure}
\begin{figure*}
  \centering
	\includegraphics[width=0.9\textwidth]{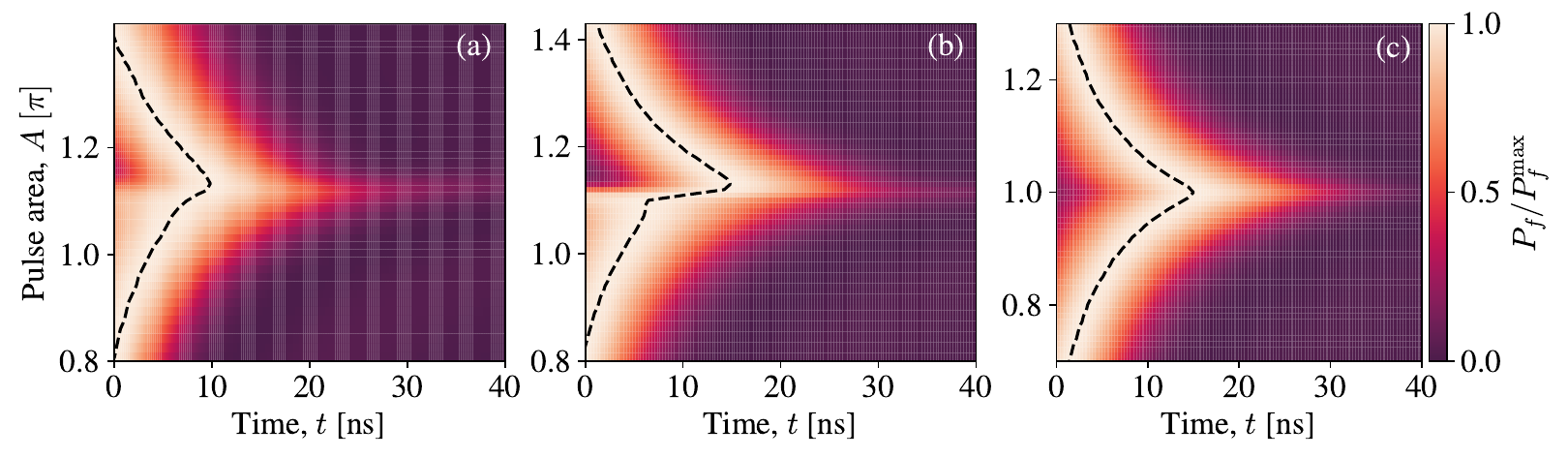}
\caption{Comparison of different model predictions for the dynamics of the power emitted into the forward-propagating mode, $P_f$, as a function of the pulse area seen by the first atom, $A$. Each time trace is normalized by its peak value, $P_f^\text{max}$. (a) We assume both fluctuating coupling strengths and a finite excitation pulse duration. (b) We assume constant coupling strengths and a finite excitation pulse duration. (c) We assume constant coupling strengths and perfect state preparation in $\ket{\psi}_\text{ideal}$.}
\label{fig:power_vs_A}
\end{figure*}

Since the atoms couple to the evanescent field of the nanofiber-guided mode, the coupling strength of each atom depends on its radial distance from the nanofiber surface. Due to the finite temperature of the atoms in the traps, this radial distance, and consequently, the atom-waveguide coupling strength, $\beta_f$, fluctuates. Moreover, the 400 excitation pulses sent per experimental sequence slightly heat the atoms during the course of the measurements. We note that, while the coupling strength fluctuates from one experimental realization to the other, one can assume that the coupling strengths do not change during the collective dynamics observed in our manuscript. This is due to the large difference of timescales between the excited state lifetime of 30~ns and the motional period of the trapped atoms of about 10~$\upmu$s. 
In order to model these fluctuations of the coupling strength in our numerical simulation, we assume a Gaussian probability distribution of $\beta_f$ values, $p(\beta_f)$. In order to ensure that $\beta_f$ only takes positive values, we truncate the distribution at $\beta_f=0$ and $\beta_f=1$. We then draw a random $\beta_f$ for each atom from this probability distribution, which is shown in Fig.~\ref{fig:dist_beta}, and propagate the light field through the ensemble of atoms using our cascaded interaction model.
We repeat this process 100 times and average the resulting model predictions. We then fit the resulting model predictions to the experimental data shown in Fig.~3 of the main manuscript and obtain a fitted mean and standard deviation of $\bar\beta_f = 0.0112$ and $\sigma_{\beta_f}=0.0065$ for the truncated Gaussian distribution, respectively. This is almost identical to the fitted values reported in our recent publication~\cite{liedl2023collective}. 

\subsection{Influence of imperfect state preparation}
Ideally, we want to initialize each atom in a coherent superposition between ground and excited state:
\begin{equation}
\ket{\psi}_\text{ideal} = \cos\left(\frac{A}{2}\right)\ket{g}-\imu\sin\left(\frac{A}{2}\right)\ket{e}.
\label{eq:psi_ideal}
\end{equation}
However, both the temperature-induced fluctuations of the coupling strength, $\beta_f$, and the finite excitation pulse length limit the fidelity of the state preparation in our experiment. In order to study the influence of these imperfections on the collective dynamics, we compare different model predictions for the dynamics of the power emitted into the forward direction, $P_f$, as a function of pulse area, $A$. 
Figure~\ref{fig:power_vs_A} shows the model predictions for 1000 atoms for three different assumptions. In (a), we include both the temperature-induced fluctuations of the coupling strengths and the finite excitation pulse duration. This is the model prediction which we also show in the manuscript. In (b), we assume a perfectly homogeneous coupling strength. Qualitatively, the dynamics is still very similar to the one displayed in (a). In both cases, the maximal delay, $t_D$, is observed for a pulse area slightly larger than $A$, and the dynamics is asymmetric with respect to maximal inversion.
However, the asymmetry is more pronounced in (b), and the maximal delay is larger. In (c), we assume that all atoms are initially prepared in $\ket{\psi}_\text{ideal}$, and that the coupling strength is constant, i.e. we neglect all experimental imperfections. 
There, the asymmetry with respect to full inversion is lifted, and the maximal delay occurs at a pulse area of $\pi$. This suggests that both the asymmetry and the fact that we observe maximal inversion for a pulse area larger than $\pi$ are due to the finite pulse duration. 
We define $A$ as the pulse area seen by the first atom of the ensemble, which is slightly larger than the average pulse area seen by the ensemble due to absorption of the excitation pulse along the ensemble. This explains why we observe maximal inversion for a pulse area that is larger than $\pi$~\cite{liedl2023collective}.
The asymmetry of the time traces with respect to full inversion is due to a combination of this pulse absorption and the fact that we work with a cascaded system. There, the first atom will always emit independently, while the last atom is driven by the light emitted by the entire ensemble. For a pulse area slightly larger than $\pi$, the last atom is prepared closer to full inversion than the first atom, and, consequently, the superradiant burst is more pronounced than for a pulse area slightly smaller than $\pi$.
\end{document}